\documentclass[conference]{IEEEtran}
\usepackage[square,numbers]{natbib}
\usepackage{cite}
\usepackage{amsmath,amssymb,amsfonts}
\usepackage{algorithmic}
\usepackage{graphicx}
\usepackage{textcomp}
\usepackage{xcolor}
\def\BibTeX{{\rm B\kern-.05em{\sc i\kern-.025em b}\kern-.08em T\kern-.1667em\lower.7ex\hbox{E}\kern-.125emX}}

\begin{document}

\title{Faster Content Delivery using RSU Caching and Vehicular Pre-caching in Vehicular Networks}

\author{
\IEEEauthorblockN{Ronella Sheryl Pereira\textsuperscript{1}, Dr. Lin Guan\textsuperscript{2}, Mao Ye\textsuperscript{3}, Ziyang Zhang\textsuperscript{4}}
\IEEEauthorblockA{
\textit{Department of Computer Science, Loughborough University}\\
Loughborough, UK, LE113TU\\
\textsuperscript{1}ronella.pereira@yahoo.in
\textsuperscript{2}l.guan@lboro.ac.uk
\textsuperscript{3}m.ye@lboro.ac.uk
\textsuperscript{4}z.zhang@lboro.ac.uk
}
}

\maketitle

%abstract page
\begin{abstract}
Most non-safety applications deployed in Vehicular Ad-hoc Network (VANET) use vehicle-to-infrastructure (V2I) and I2V communications to receive various forms of content such as periodic traffic updates, advertisements from adjacent road-side units (RSUs). In case of heavy traffic on highways and urban areas, content delivery time (CDT) can be significantly affected. Increase in CDT can be attributed to high load on the RSU or high volume of broadcasted content which can flood the network. Therefore, this paper suggests a novel caching strategy to improve CDT in high traffic areas and three major contributions have been made: (1) Design and simulation of a caching strategy to decrease the average content delivery time; (2) Evaluation and comparison of caching performance in both urban scenario and highway scenario; (3) Evaluation and comparison of caching performance in single RSU and multiple RSUs. The simulation results show that caching effectively reduces the CDT by 50\% in urban scenario and 60-70\% in highway scenario. 
\end{abstract}

%\begin{IEEEkeywords}
%Vehicular Network, RSU Caching, Content Caching, Vehicular pre-caching
%\end{IEEEkeywords}

%intro page
\section{Introduction}
\label{section:intro}

Vehicular Ad-Hoc Network (VANET) is an emerging technology that has gathered incremental attention in both academia and industry in recent years. It is a key component of Intelligent Transportation System (ITS) framework in which mobile vehicles forming nodes can communicate cooperatively and wirelessly \citep{zekri2018heterogeneous}. VANET can be used to establish communication between vehicles or vehicles and nearby road-side equipment \citep{liang2015vehicular}. Various safety and non-safety applications control transport, traffic management, collision  avoidance, entertainment  and  commercial  services \citep{zhou2020evolutionary} that use V2I for communication.

\subsection{Background and Motivation}
One of the main challenges in VANET is connectivity \citep{al2014comprehensive}. Good performance in vehicular communication depends on the connectivity and hence it is a key issue for VANET under constraints where vehicles are moving with high speed, geographical restrictions, dynamic network topology, limited channel bandwidths \citep{dixit2016vanet}. The dynamic network topology in VANET causes inter-vehicular links to connect and disconnect frequently \citep{lee2021vanet, liang2015vehicular}. This often causes a delay in receiving traffic content from RSU or unnecessary repeated broadcasts. This paper suggests caching embedded VANET in order to overcome this problem.  

Various caching methods have been proposed in previous works. The caching at RSUs wherein popular contents are stored in the RSUs caches, allows vehicles to download content in 70\% lesser time \citep{ding2015roadside}. In the paper \citep{ding2015roadside}, the simulation has only considered low traffic load of few vehicles. Another paper \citep{tian2015lce} considers the low delivery ratio in urban environments from RSU to vehicles. It identifies obstacles (e.g., buildings) as the main issue, suggests lesser dependency on V2I and thus proposes the vehicular caching in vehicle-to-vehicle (V2V) as an alternative. However, it fails to suggest a strategy to improve the V2I communications and the impact of high traffic loads on RSUs. This paper considers both the urban and highway scenarios with heavy traffic and suggests the RSU caching integrated with vehicular pre-caching to improve CDT.

The paper is organised as follows: Section 2 describes the methodology used to develop the system. The various modelling and implementation methods to achieve this goal is elaborated. Section 3 shows the setup of the simulation,
the experiments conducted on the system model and metrics used for analysis. Section 4 summarises the results of the various simulations conducted and gives an in-depth analysis. Section 5 describes the overall conclusions and future works.

%proj design page
 
\section{Caching Methodology}
\label{section:design}

The initial step to designing the cache, was to target the appropriate applications that would benefit from caching. According to the authors, Glass et al. caching may not be suitable for time-critical applications (e.g., collision avoidance) due to the fact that caching stores data for a certain amount of time \citep{glass2017leveraging}. So the caching is suitable for applications that retain data for a longer period of time such as location based services, traffic information or infotainment applications.

The cache is designed using the Information-Centric Network (ICN) naming convention, where a content is identified using the content name rather than IP address. Using hierarchical content name for example '/traffic/geolocation/timestamp/datatype' where '/traffic' represents the application \citep{amadeo2016information}. In our design, the contents are cached using the content name as the key that maps the content name to the appropriate requested content.

The cache uses the Least Recently Used (LRU) method for cache replacement policy. When the cache is full, LRU policy replaces the content in the cache that has not been accessed for the longest time with the new content \citep{khelifi2020network}. But in most cases the cache will not require any replacement policy as the vehicular caches are assumed to be infinite memory and the RSU also has enough storage capacity to handle small to medium sized content text files. However, such assumptions cannot be allowed while designing any system and hence the LRU replacement policy is the most suitable  to be used. 

The unique feature of VANET wherein the vehicles are restricted by a certain mobility pattern with fixed road layouts makes vehicular caching considerably feasible and can be leveraged to our benefit. For instance, a large group of vehicles travelling on the highway are always headed towards the same direction, travelling in a platoon and hence may have similar content requests sent to the RSU \citep{glass2017leveraging}. Moreover, if a content has been requested by one vehicle, there exists a high probability that it will be requested by the neighbouring node in a nearby time frame. Therefore, for the placement policy, the caching is implemented at RSU and the content broadcasted is pre-cached at every node within the communication range of the RSU in the case of vehicular caching.

The V2I/I2V communications are based on the broadcast protocol using DSRC standard (IEEE 802.11p). According to Din et al., data is disseminated throughout the network by two basic mechanisms: push-based and pull-based method. In pull-based mechanism, a node requests desired data or content, and the node having the expected content will respond back with the corresponding data packet \citep{din2020left}. This is widely used in infotainment applications or in cases where a vehicle requests for traffic data or road conditions. Hence, the vehicle to RSU communication is designed to be pull-based. The authors Yi et al., suggest that a mobile edge server (MES) is usually deployed alongside RSU. The MES is capable of collecting and processing the information of various road conditions and traffic accidents \citep{yi2021content}. Additionally, MES can also have any popular contents like nearby restaurant, hotel, tourist spots information stored in it. The vehicles don't have direct access to the server but can access the contents through a nearby RSU.

%proj imp page
 
\section{Simulation Setup}
\label{section:setup}
Simulation is the depiction of the real world operations in a fixed time frame. In the case of VANET, testing research models on actual vehicles and roads can be risky as any failed test could have serious consequences. A proper simulator that imitates the real world as closely as possible, is required. In this paper, an unified framework federating Simulation of Urban Mobility (SUMO), OMNET++ and Vehicle in Network Simulation (VEINS) framework has been used to evaluate the performance of the proposed caching system. The common parameters set for the simulations are given in Table 1.
\begin{table}[htb]
 \caption{Parameters of the simulation}
 \begin{center}
 \begin{tabular}{|c|c|}
 \hline
 \textbf{Parameters} & \textbf{Value of the parameters}\\
 \hline
 Node Acceleration & 2.6ms$^{-2}$ \\
  \hline
  Node Deceleration & 4.5ms$^{-2}$ \\
  \hline
  Maximum Node Speed & 14ms$^{-1}$ \\
  \hline
  Minimum Node Gap & 2.5m \\
  \hline
 Node beacon Interval & 10s \\
 \hline 
 Header Length & 80bit \\
 \hline
 Transmitter Power & 20mW \\
 \hline
 Bit Rate & 6Mbps \\
  \hline
 Noise floor & -98dBm \\
 \hline
 Minimum Power Level & -110dBm \\
 \hline
 Antenna Height & 1.895m \\
  \hline
  Decider Center Frequency & 5.89 GHz \\
  \hline
\end{tabular}
\end{center}
\end{table}

%proj imp page
 \subsection{Scenario Description}
The data used for the experiments is static and generated only on the server applications and each generated content file has a unique content name. The content names start from '/traffic/1' to '/traffic/10'. Each vehicle will randomly request for content from these ten content files. The request is sent every 10 seconds until the vehicle receives the requested content. The performance is analysed based on two scenarios: Urban and Highway which are discussed below: 
\\
\\\textit{Urban Scenario}: 
\begin{figure}[!htb]
    \centerline{
    \includegraphics[width=0.5\textwidth,height=0.6\textheight,keepaspectratio]{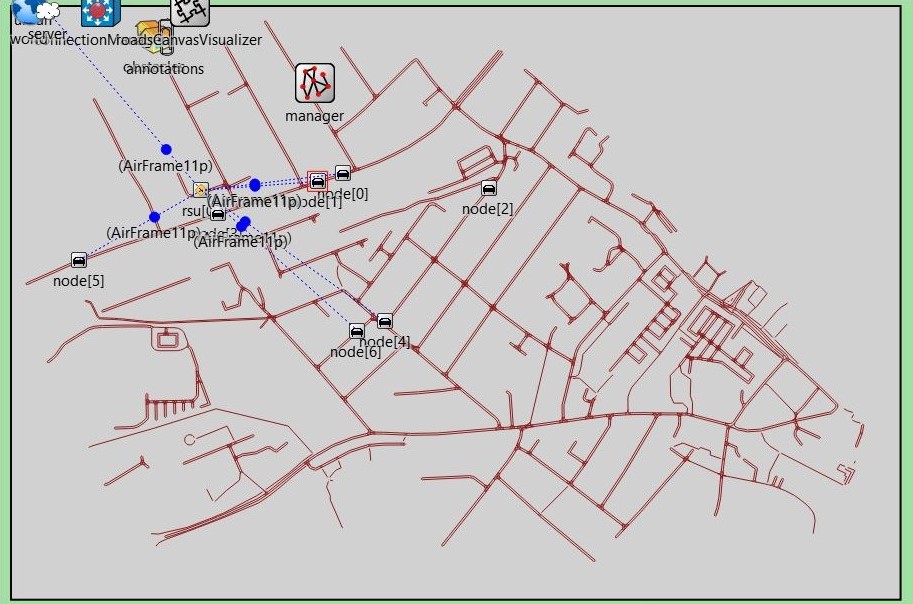}}
    \caption{Simulation of Urban Scenario-Single RSU}
    \label{urban3}
\end{figure}
For single RSU as seen in Figure \ref{urban3}, two roads are under a single RSU having a coverage of 400m. Since it is an urban area, the traffic flow will be irregular i.e., vehicles will be entering and exiting these two routes at random time intervals. Traffic load of 20 vehicles in 144s and 40 vehicles in 230s are considered. 

For multiple RSU, the two roads of the urban scenario are divided between two different RSUs, each having a coverage of 270m. In this experiment, the two RSUs are placed in such a way that they do not have overlapping areas. Each RSU is separately connected to the server. Traffic load of 40 vehicles in 230s and 60 vehicles in 430s are considered.
\\
\\\textit{Highway Scenario}: 
\begin{figure}[!htb]
    \centerline{
    \includegraphics[width=250pt,height=200pt]{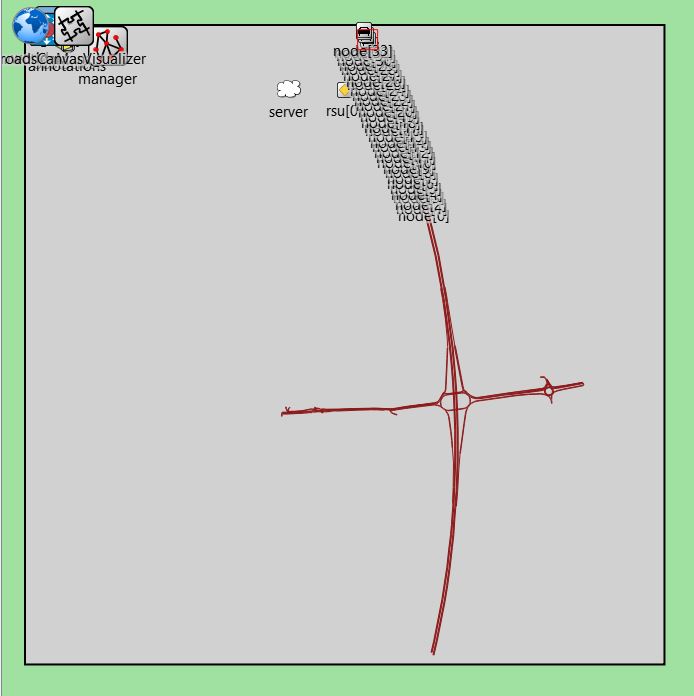}}
    \caption{Simulation of Highway Scenario-Single RSU}
    \label{highway3}
\end{figure}
In single RSU experiment as seen in Figure \ref{highway3}, the highway has an adjacent RSU having coverage of 400m. Since it is a highway, the traffic will be in a flow or platoon of vehicles all headed in the same direction. High traffic load of 300 vehicles in which each vehicle enters the simulation per second is selected. 

For the next experiment, there are three RSUs overlapping each other and only one RSU is connected to the server. The two RSUs are used as relay RSUs and all three RSUs are placed along the highway. The coverage area of each RSU is 200m. High traffic load is also considered here, same as single RSU.

\subsection{Performance Metrics}
\begin{itemize}
\item \textit{Average Content Delivery Time}: Content delivery time (CDT) is the time taken by a vehicle to receive the requested content since it first sent a request to the nearby RSU. Average CDT is the average of all delivery times completed at a given point in time.
\item \textit{Number of requests to server}: This denotes the number of requests sent to the edge server and usually depicts the usage of internet services.
\item \textit{Number of requests to the RSU}: This denotes the number of requests sent to the RSU and usually depicts the load on infrastructural facilities.
\item \textit{Cache Hit Ratio (CHR)}: A cache hit means that the request for content is found in the cache and a cache miss occurs if the content is not found in the cache. Therefore, CHR is defined as the ratio between the number of cache hits to the sum of the number of cache hits and cache misses \citep{zhang2015survey}.
\end{itemize}

\section{Results and Analysis}
\label{section:results}

\subsection{Urban Scenario}
\subsubsection{Single RSU without caching}
\begin{figure}[!htb]
    \centerline{\includegraphics[width=0.5\textwidth,height=0.6\textheight,keepaspectratio]{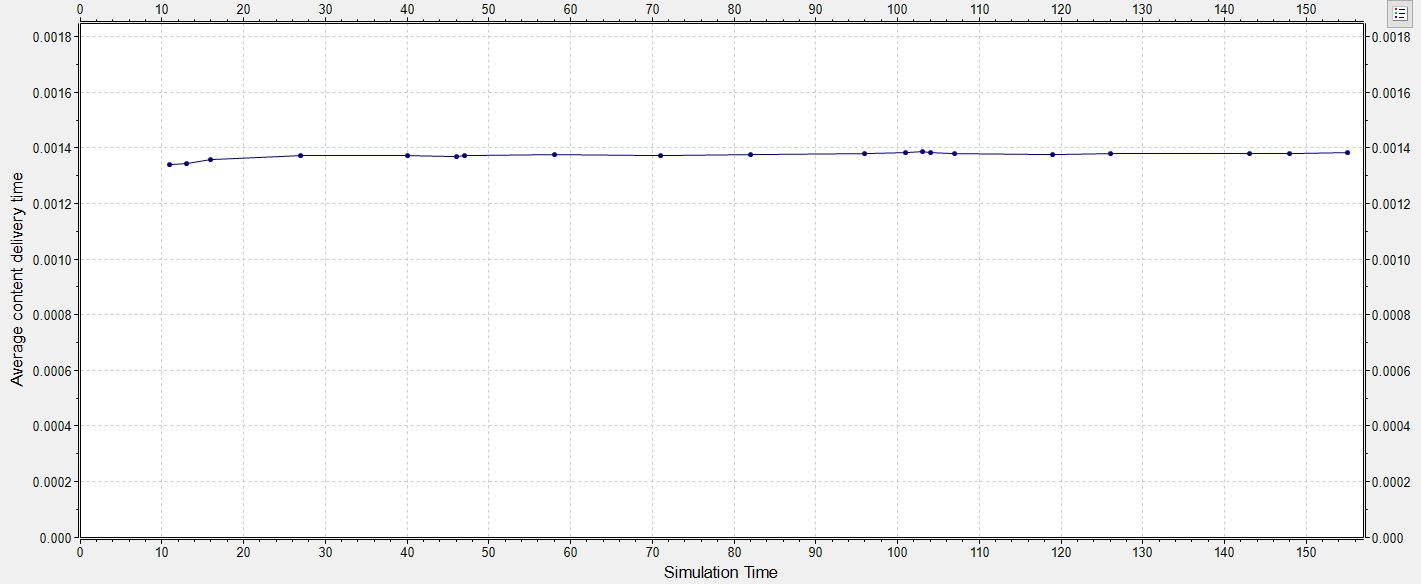}}
    \caption{Average CDT for 20 vehicles - No caching}
    \label{unc1}
\end{figure}

Figure \ref{unc1} is the graph for average CDT when we consider 20 vehicles. It shows that each vehicle receives its requested content within 1.4 milliseconds without caching.

\begin{figure}[!htb]
    \centerline{
    \includegraphics[width=0.5\textwidth,height=0.6\textheight,keepaspectratio]{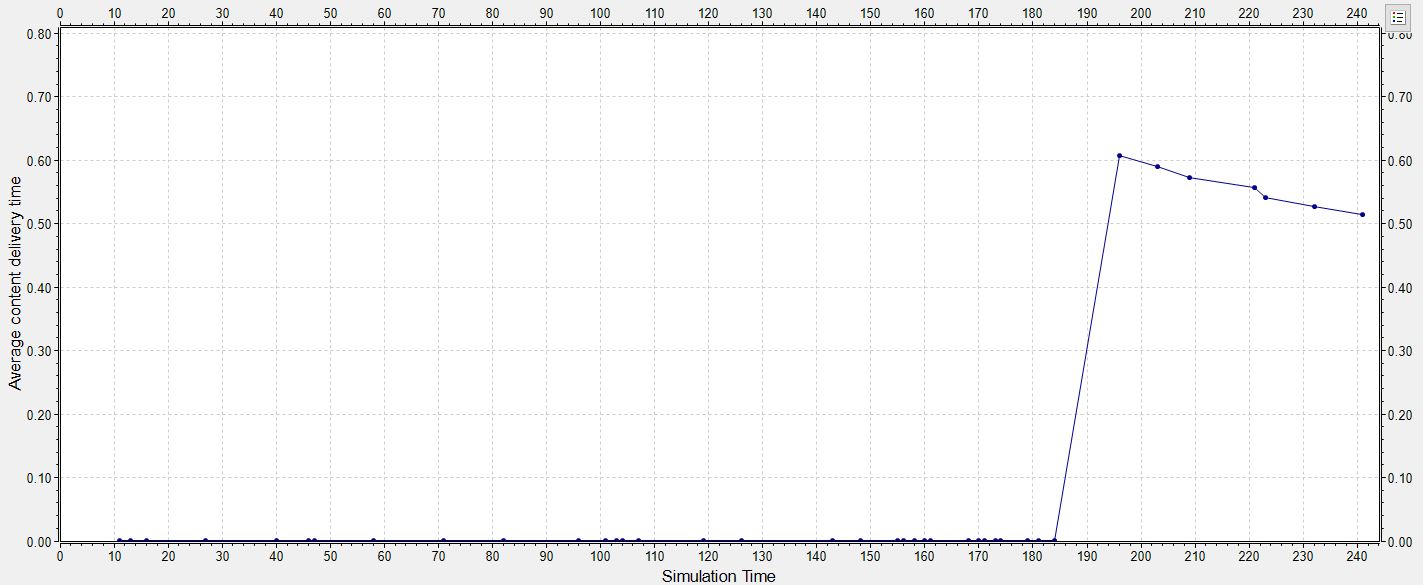}}
    \caption{Average CDT for 40 vehicles - No caching}
    \label{unc2}
\end{figure}

When we increase the vehicles from 20 to 40, we can see in the Figure \ref{unc2} that there is a point in simulation time where there is a sudden rise in the graph. This is due to the fact that the increase in traffic leads to the increase in beacons being sent, which causes delay in some vehicles receiving their requested content. 

\begin{figure}[!htb]
    \centerline{
    \includegraphics[width=0.5\textwidth,height=0.6\textheight,keepaspectratio]{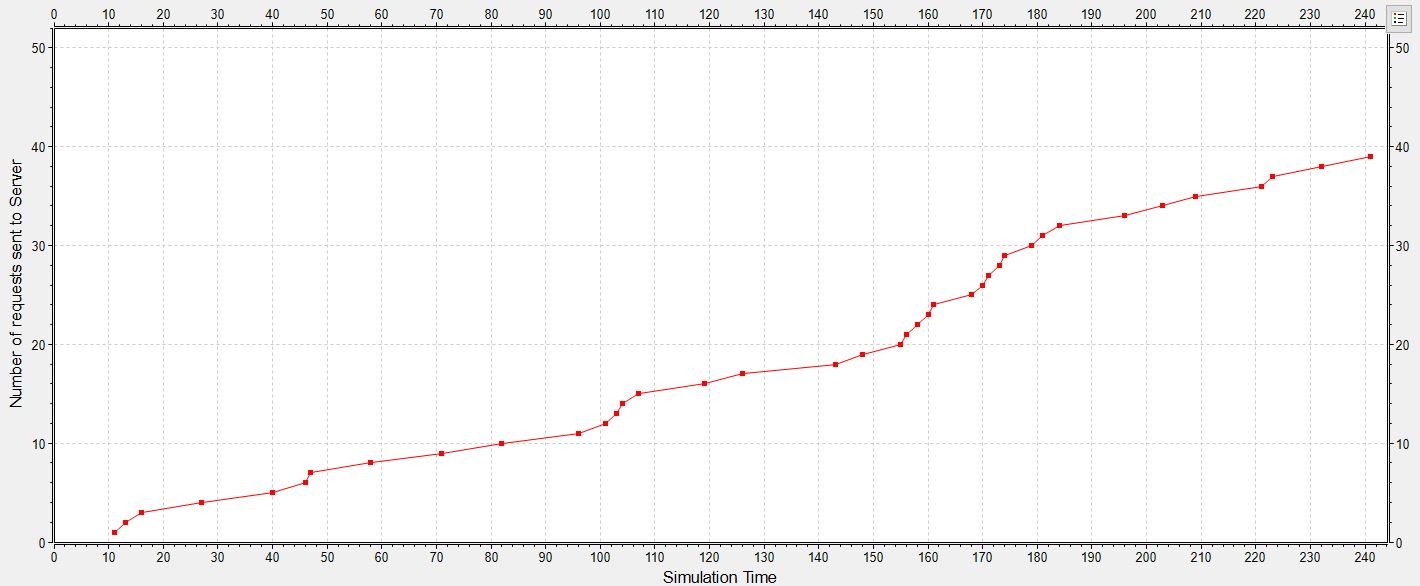}}
    \caption{Number of requests sent to server for 40 vehicles}
    \label{uns2}
\end{figure}

\begin{figure}[!htb]
    \centerline{
    \includegraphics[width=0.5\textwidth,height=0.6\textheight,keepaspectratio]{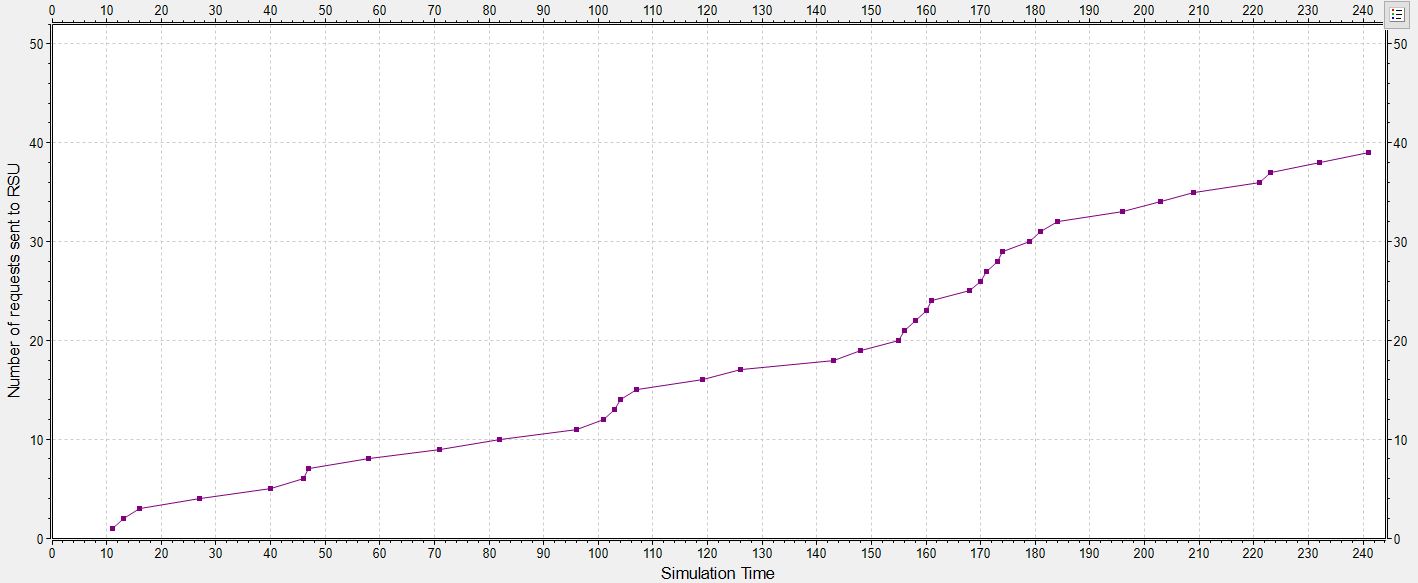}}
    \caption{Number of requests sent to RSU for 40 vehicles}
    \label{unr2}
\end{figure}

The Figures \ref{uns2} and \ref{unr2} show the number of requests sent to the server and RSU for 40 vehicles. This depicts that the number of vehicles is equal to the number of content requests sent to RSU which are forwarded to the server. The increase in the number of vehicles means increased load on RSU and subsequently the server.

\subsubsection{Single RSU with caching strategy}
\begin{figure}[!htb]
    \centerline{
    \includegraphics[width=0.5\textwidth,height=0.6\textheight,keepaspectratio]{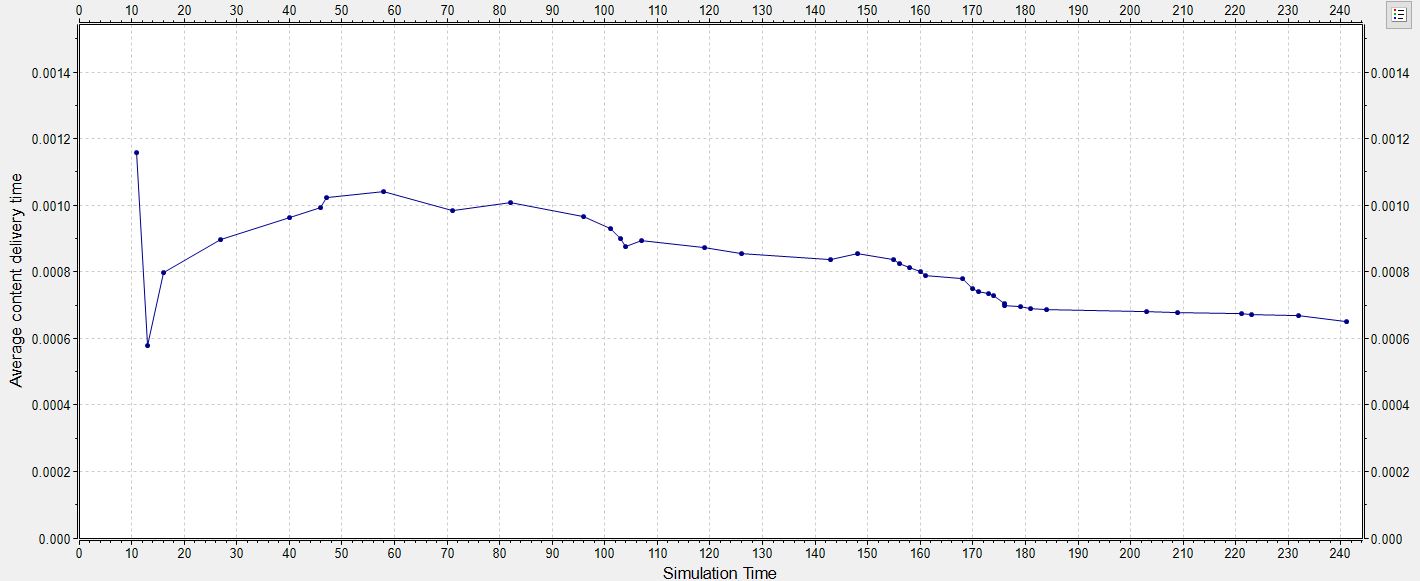}}
    \caption{Average CDT for 40 vehicles - Single RSU}
    \label{urvc2}
\end{figure}

 In the Figure \ref{urvc2}, there is a huge improvement in CDT as compared to having no caching. The sudden spike seen in a high traffic load of 40 vehicles disappears due to the available cached contents at RSU and vehicles. The RSU is less occupied with sending and receiving redundant content copies from the server. 

\begin{figure}[!htb]
    \centerline{
    \includegraphics[width=0.5\textwidth,height=0.6\textheight,keepaspectratio]{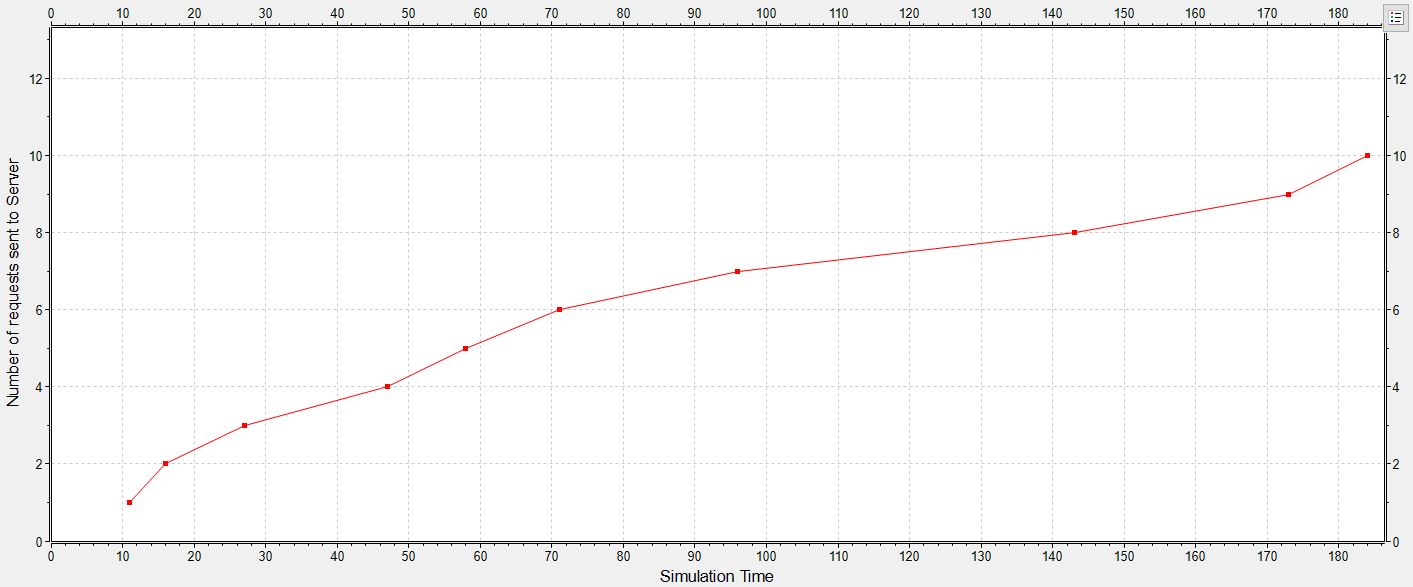}}
    \caption{Number of request sent to server for 40 vehicles}
    \label{urs2}
\end{figure}

As seen in the Figure \ref{urs2}, the number of forwarded requests goes down to 10 which is the number of unique content files. This implies reduction in internet usage in the VANET as lesser requests are sent from RSU to server. 

\begin{figure}[!htb]
    \centerline{
    \includegraphics[width=0.5\textwidth,height=0.6\textheight,keepaspectratio]{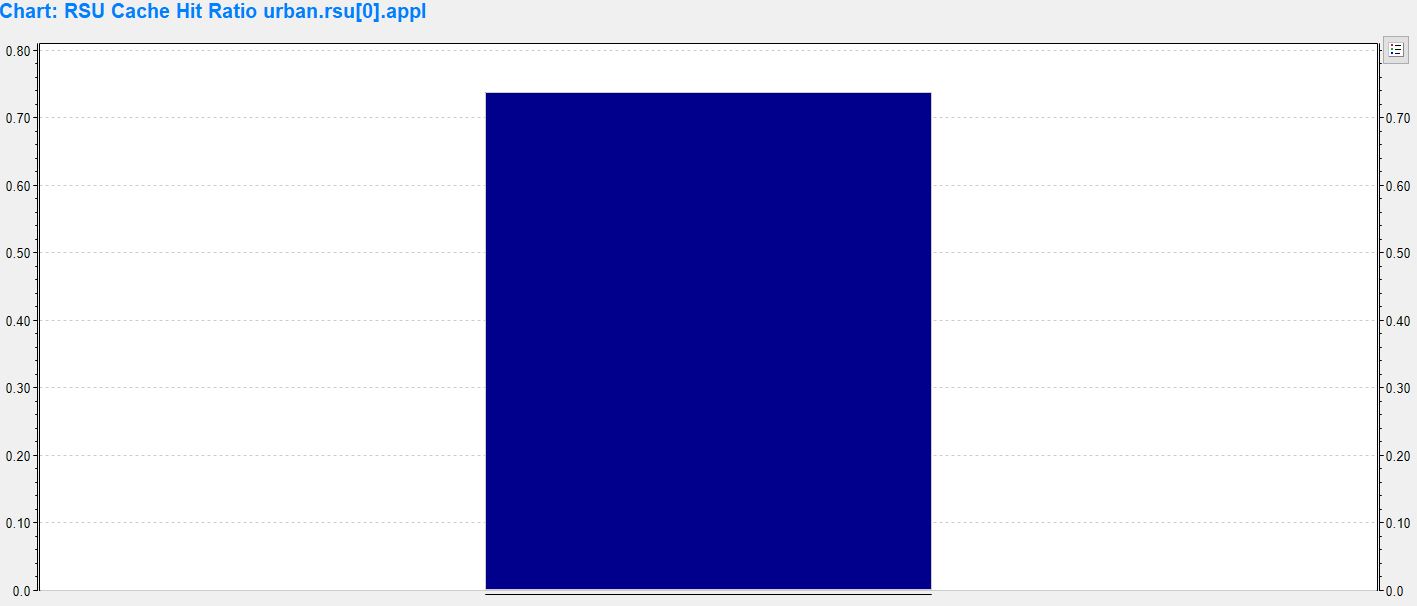}}
    \caption{CHR for 40 vehicles - Single RSU}
    \label{a2}
\end{figure}

Figure \ref{a2} shows a CHR of 0.70 depicting 70\% cache hits.

\subsubsection{Multiple RSUs with caching strategy}
\begin{figure}[!htb]
    \centerline{
    \includegraphics[width=0.5\textwidth,height=0.6\textheight,keepaspectratio]{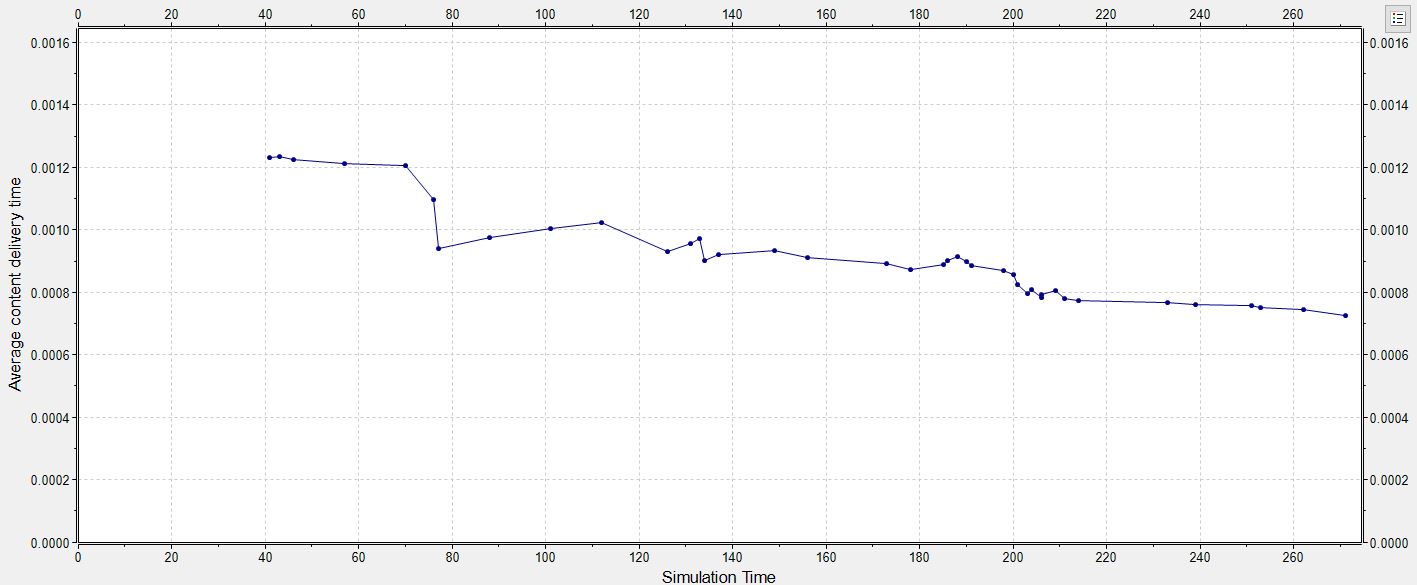}}
    \caption{Average CDT for 40 vehicles - Multiple RSUs}
    \label{c2}
\end{figure}

\begin{figure}[!htb]
    \centerline{
    \includegraphics[width=0.5\textwidth,height=0.6\textheight,keepaspectratio]{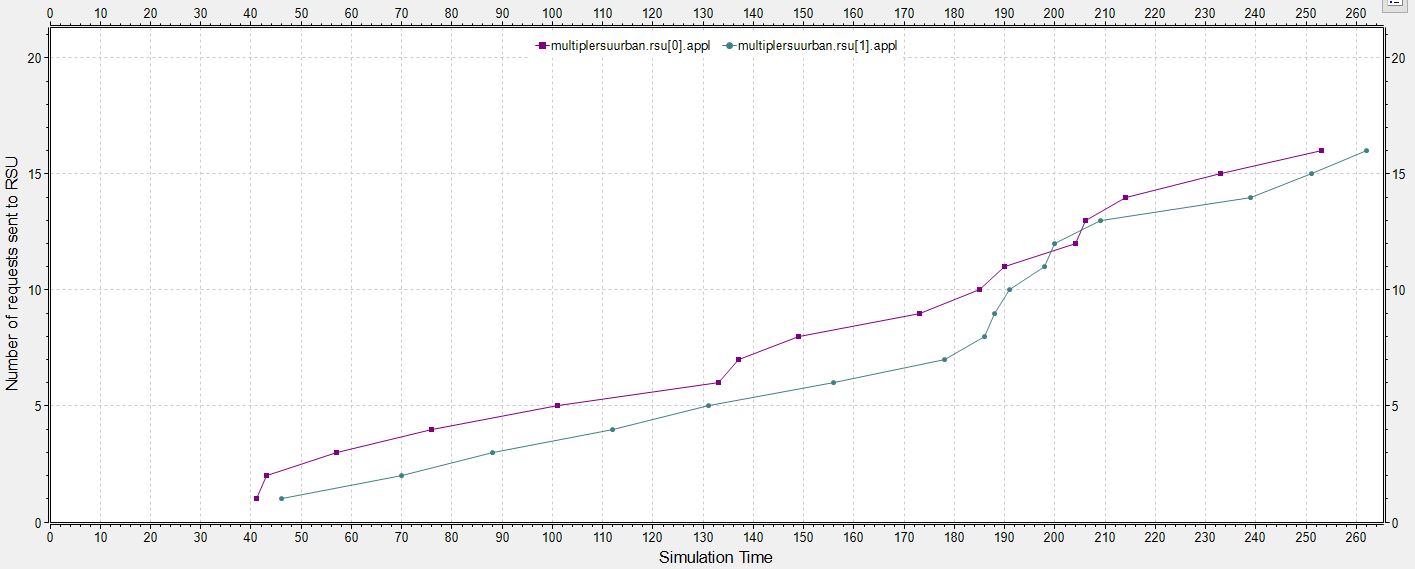}}
    \caption{Number of requests sent to RSU for 40 vehicles}
    \label{r2}
\end{figure}

Figure \ref{c2} shows that multiple RSUs performs the same as single RSU in terms of delivery time.
However, as seen in Figure \ref{r2}, the number of requests handled by the RSUs can be divided among them. Thus, reducing the load on a single RSU.

\begin{figure}[!htb]
    \centerline{
    \includegraphics[width=0.5\textwidth,height=0.6\textheight,keepaspectratio]{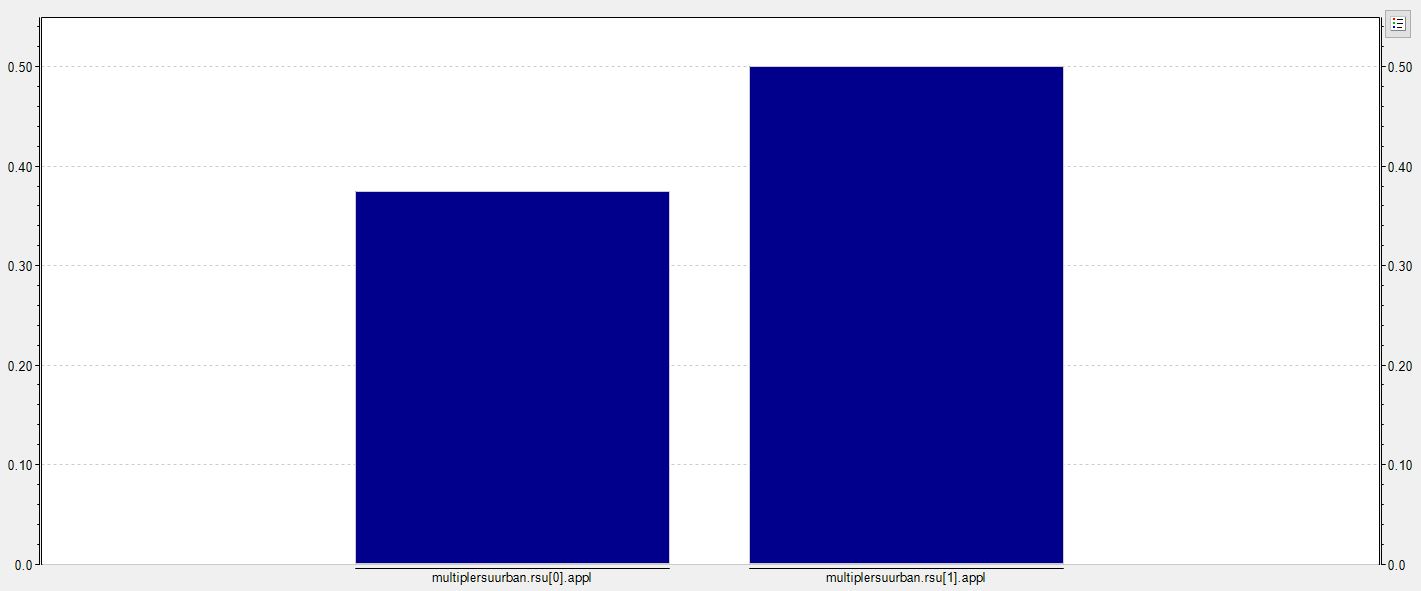}}
    \caption{CHR for 40 vehicles - Multiple RSUs}
    \label{ma2}
\end{figure}

\begin{figure}[!htb]
    \centerline{
    \includegraphics[width=0.5\textwidth,height=0.6\textheight,keepaspectratio]{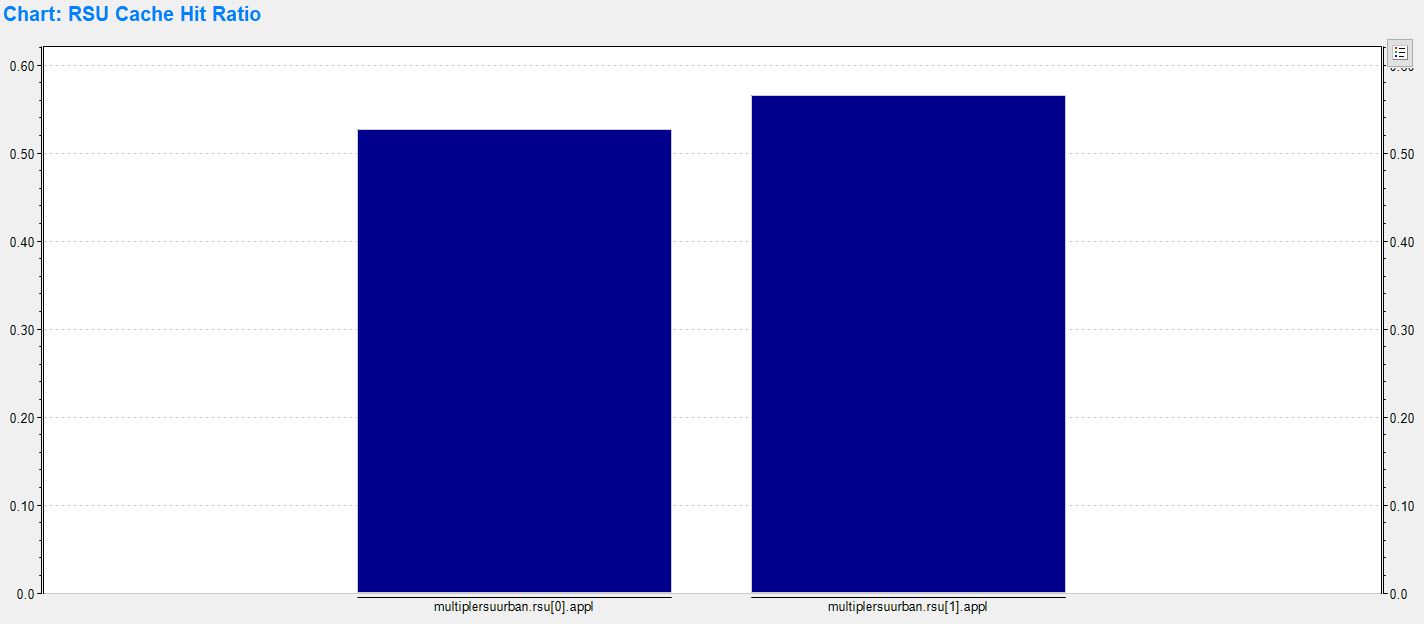}}
    \caption{CHR for 60 vehicles - Multiple RSUs}
    \label{ma3}
\end{figure}

Figure \ref{ma2} and \ref{ma3} show that the CHR is decreased in comparison with single RSU but it increases with increase in traffic load. This is because the two RSU caches are independant of each other.

\subsection{Highway Scenario}
\subsubsection{Single RSU without caching}
\begin{figure}[!htb]
    \centerline{
    \includegraphics[width=0.5\textwidth,height=0.6\textheight,keepaspectratio]{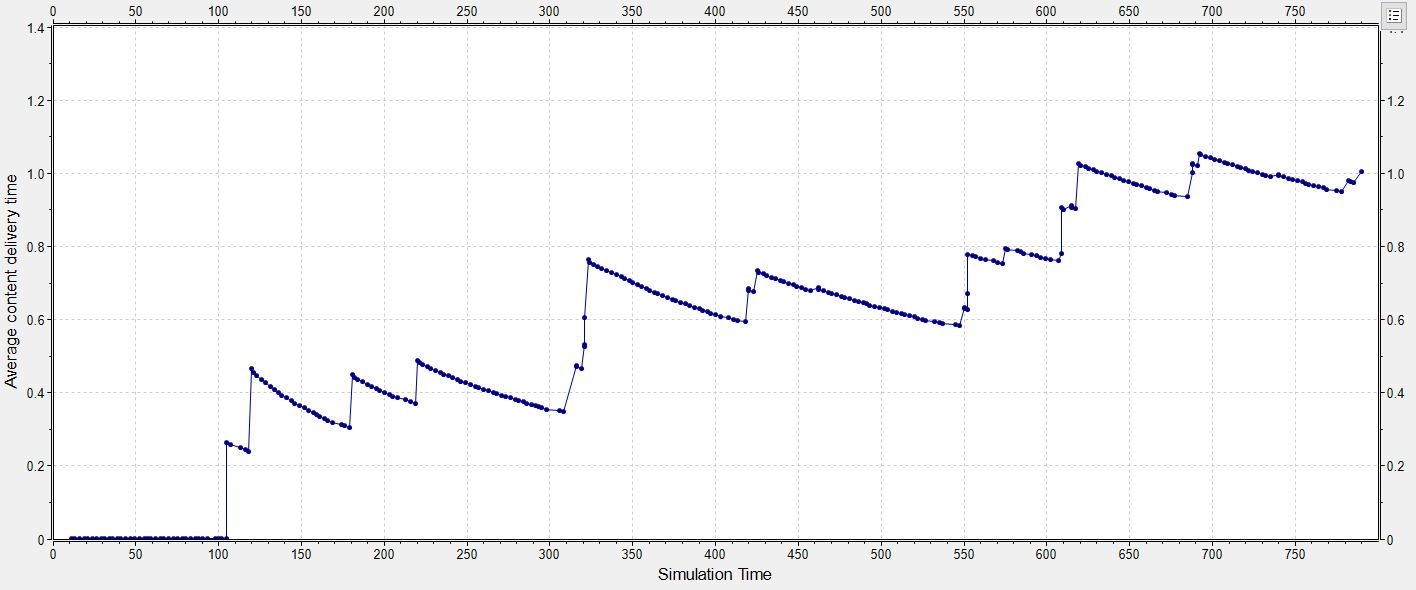}}
    \caption{Average CDT for 300 vehicles - No caching}
    \label{hnc2}
\end{figure}

Figure \ref{hnc2} shows the average CDT for 300 vehicles without caching. There are waves of sudden rise in CDT till it hits one second.

\begin{figure}[!htb]
    \centerline{
    \includegraphics[width=0.5\textwidth,height=0.6\textheight,keepaspectratio]{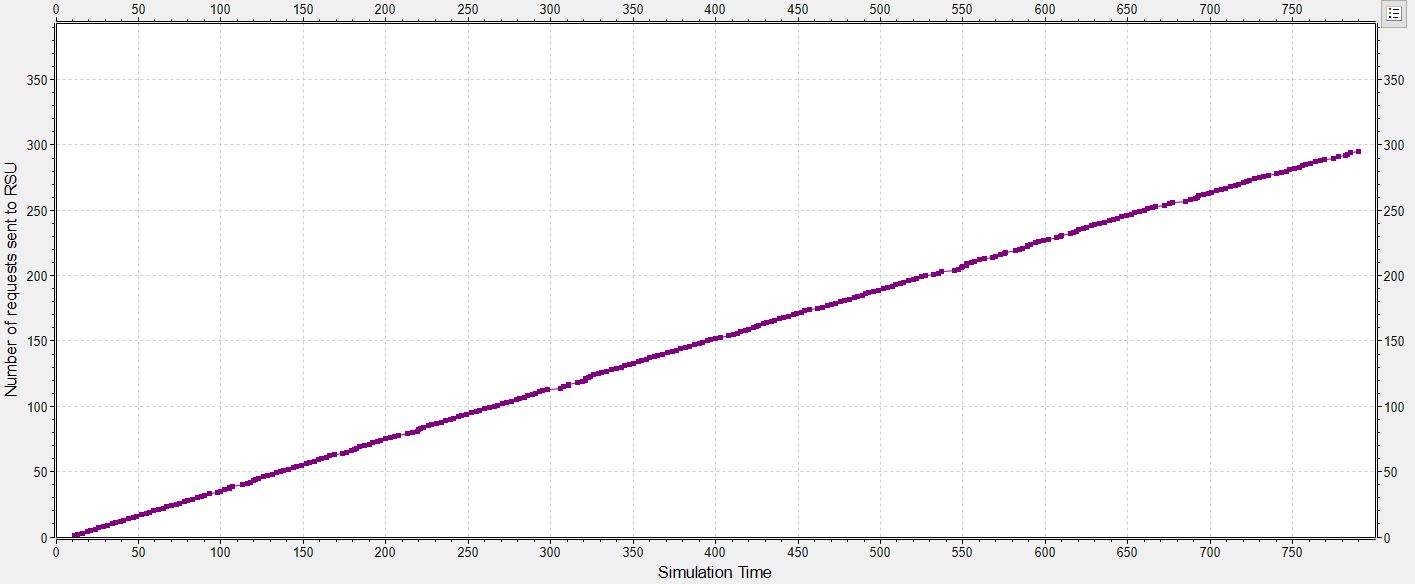}}
    \caption{Number of requests sent to RSU for 300 vehicles}
    \label{hnr2}
\end{figure}

The number of requests sent to RSU and server without caching is less than or equal to number of vehicles as seen in Figure \ref{hnr2}. The content is requested uniformly over time as the vehicles travel in a platoon as compared to the urban scenario which is irregular.

\subsubsection{Single RSU with caching strategy}
\begin{figure}[!htb]
    \centerline{
    \includegraphics[width=0.5\textwidth,height=0.6\textheight,keepaspectratio]{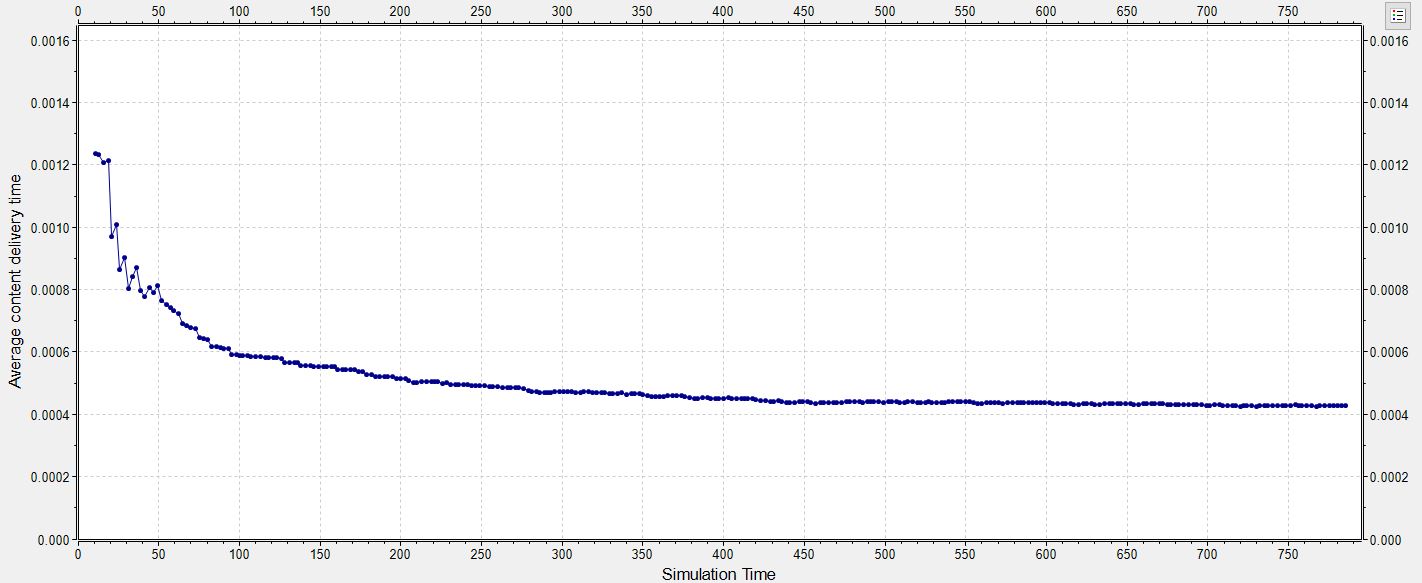}}
    \caption{Average CDT for 300 vehicles - Single RSU}
    \label{hrvc2}
\end{figure}

The average CDT has improved to nearly 0.4 milliseconds as seen in Figure \ref{hrvc2}. This could be because in highway all the vehicles are moving closely in the same direction and thus have more vehicles pre-caching the contents.

\begin{figure}[!htb]
    \centerline{
    \includegraphics[width=0.5\textwidth,height=0.6\textheight,keepaspectratio]{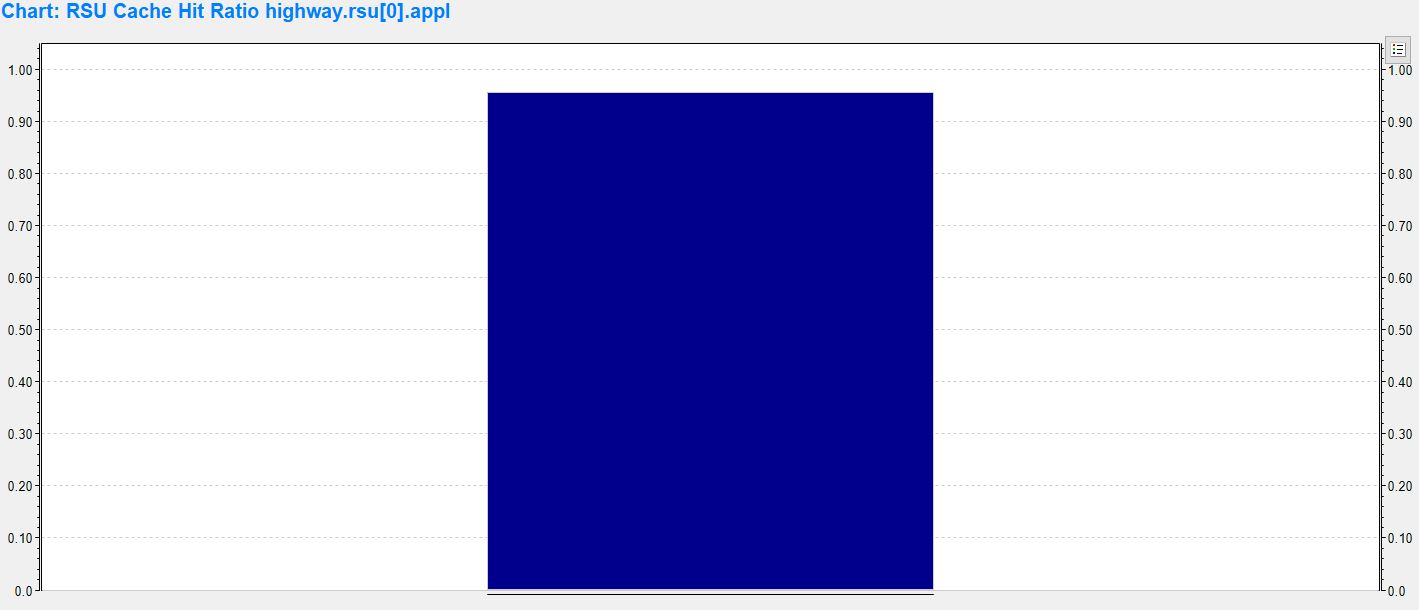}}
    \caption{CHR for 300 vehicles - Single RSU}
    \label{hrv2}
\end{figure}

\begin{figure}[!htb]
    \centerline{
    \includegraphics[width=0.5\textwidth,height=0.6\textheight,keepaspectratio]{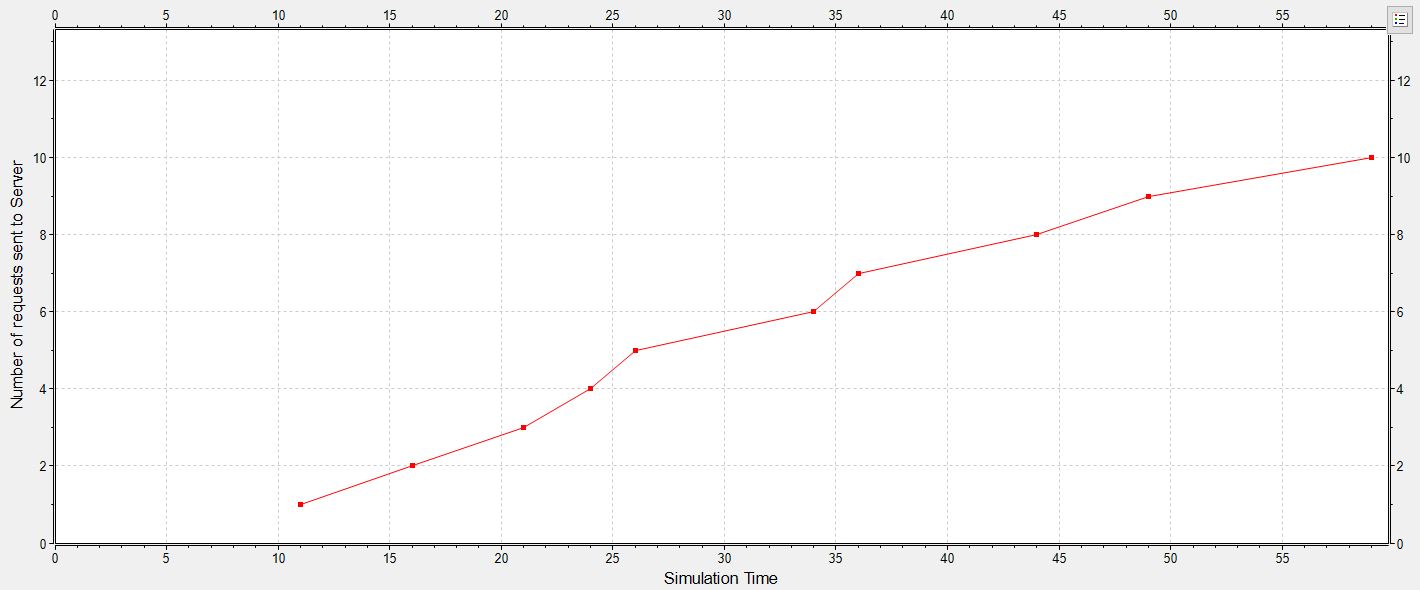}}
    \caption{Number of requests sent to server for 300 vehicles}
    \label{hrs2}
\end{figure}

\begin{figure}[!htb]
    \centerline{
    \includegraphics[width=0.5\textwidth,height=0.6\textheight,keepaspectratio]{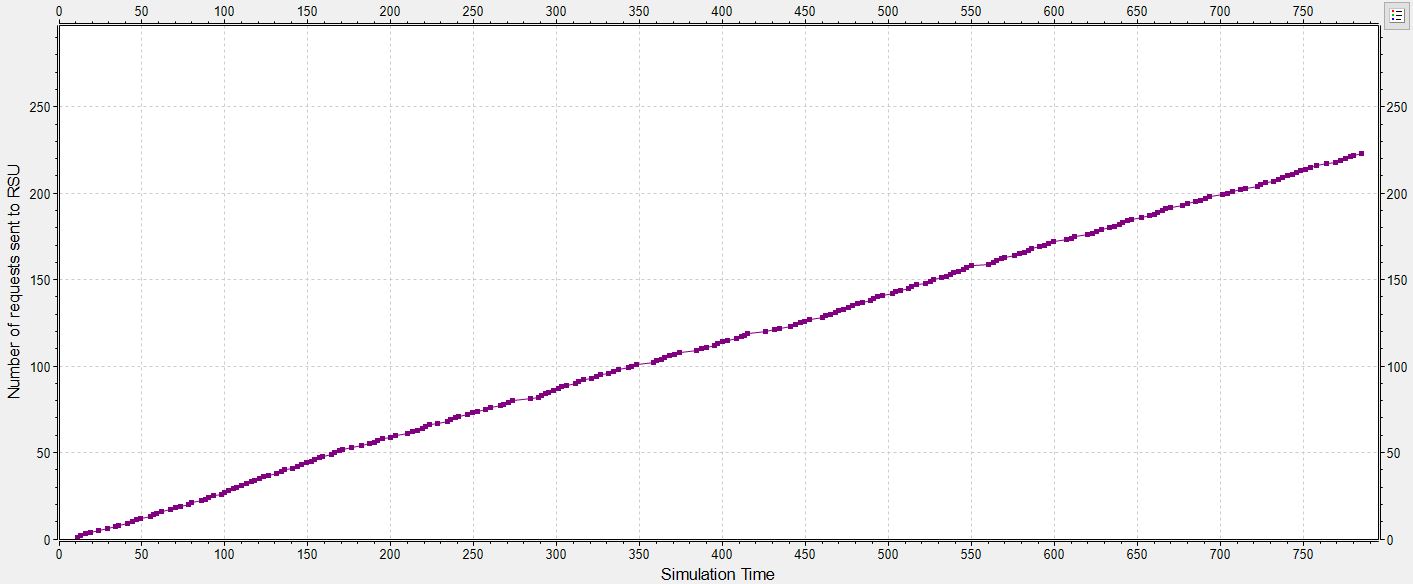}}
    \caption{Number of requests sent to RSU for 300 vehicles}
    \label{hrvr}
\end{figure}

Although the pre-caching at vehicle slightly lowers the CHR of the RSU cache, it is almost negligible as shown in Figure \ref{hrv2}.
Another impact of vehicular pre-caching is seen in the number of requests sent to server and RSU as shown in Figures \ref{hrs2} and \ref{hrvr}.
\subsubsection{Multiple RSUs with caching strategy}
\begin{figure}[!htb]
    \centerline{
    \includegraphics[width=0.5\textwidth,height=0.6\textheight,keepaspectratio]{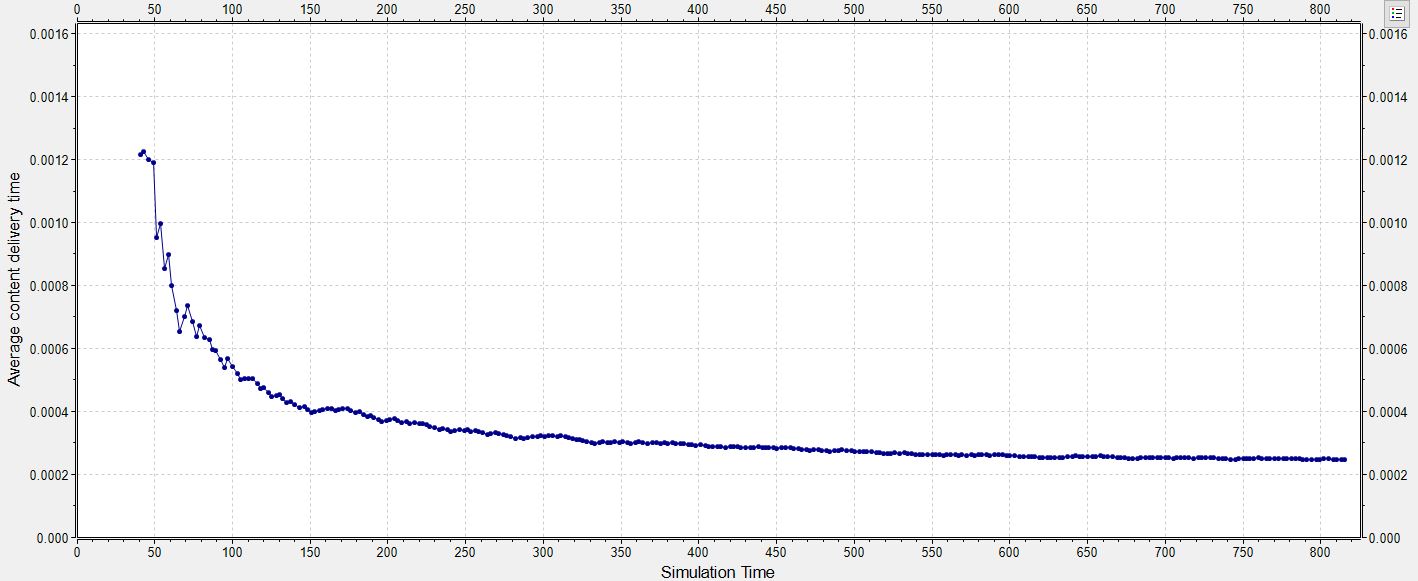}}
    \caption{Average CDT for 300 vehicles - Multiple RSUs}
    \label{hc2}
\end{figure}

With the addition of two more RSUs the average CDT has dropped down to 0.25 milliseconds as seen in Figure \ref{hc2}. The reason is that the RSUs are in communication with each other and the two RSUs broadcast the cached contents, thus increasing the number of pre-cached contents.

\begin{figure}[!htb]
    \centerline{
    \includegraphics[width=0.5\textwidth,height=0.6\textheight,keepaspectratio]{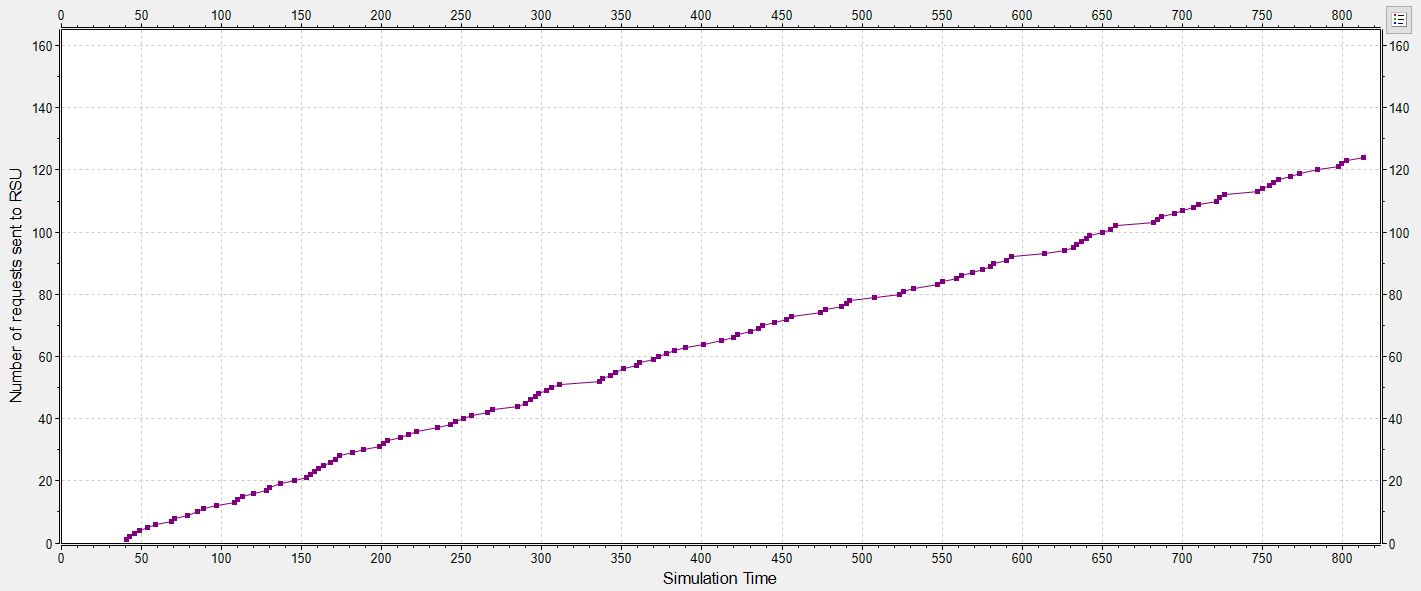}}
    \caption{Number of requests sent to RSU for 300 vehicles}
    \label{mhr2}
\end{figure}

The Figure \ref{mhr2} even shows an overall reduction in the number of requests sent to RSU. This is better than the results of urban scenario because the additional RSUs are used only for relaying cached contents.
%proj conclusions page
\section{Conclusions \& Future work}
\label{section:conclusion}
In this paper, we originally deployed a novel caching strategy in VANET. The state of art research on VANET and its identified challenges, helped in designing the caching methodology. VANET embedded with the caching system, demonstrated better results for both urban and highway scenarios as compared to no caching. 

In the case of urban scenario, it is seen that for 40 vehicles, the average content delivery over time drops from 1.2 milliseconds to 0.6 milliseconds. In the case of highway scenario, caching has significantly better results bringing the average CDT to 0.4 milliseconds. 

The effect of implementing multiple RSUs was further analysed. The results show that the placement of RSU is very crucial to obtain the best outcome of caching. Multiple RSUs that are independant from each other as in urban scenario, only improves the amount of requests each RSU can handle and thus is suitable for the large-scale scenarios with high traffic loads. Relaying cache contents in highway scenario with multiple RSUs improved the caching performance drastically. In highway scenarios, addition of multiple RSUs decreases CDT to 0.25 milliseconds and the number of requests sent to RSU are also reduced by 50\%. There are also several possibilities of future research: 
\begin{itemize}
    \item Caching can be implemented for V2V communications and this would involve adding the caching to the routing protocol. VEINS has a subproject called VEINS\_INET which can be used to implement this in VEINS.
    \item Artificial Intelligence can be used to enhance caching in VANET by taking the caching decision on which content would be cached. However, the research in this area is limited.
    \item ICN-based VANET is a useful future work as ICN architecture comes with the feature of caching. Also, the IP address based internet stack could be replaced by a more content-centric network stack in future.
\end{itemize}

%bibliography
\renewcommand\bibname{REFERENCES}
\bibliographystyle{IEEEtranN}
\bibliography{vanet}

\end{document}